\documentclass[conference]{IEEEtran}

\usepackage{graphicx}
\usepackage{amsmath}
\usepackage{color}
\usepackage{cite}
\usepackage{verbatim}
\definecolor{darkgray}{gray}{.4}

\newcommand{\eqd}{\stackrel{\triangle}{=}}

\newcommand {\bp} {\mbox{\boldmath $p$}}

\newcommand {\br} {\mbox{\boldmath $r$}}

\newcommand {\bx} {\mbox{\boldmath $x$}}

\newcommand {\bE} {\mbox{\boldmath $E$}}

\newcommand {\bP} {\mbox{\boldmath $P$}}

\newcommand {\bW} {\mbox{\boldmath $W$}}
\newcommand {\bX} {\mbox{\boldmath $X$}}

\newcommand{\calE}{{\cal E}}

\newcommand{\calI}{{\cal I}}

\newcommand{\calN}{{\cal N}}

\newcommand{\calX}{{\cal X}}

\IEEEoverridecommandlockouts

\begin{document}
\thispagestyle{empty}
\setcounter{page}{1}
\title{Physics of the Shannon Limits}

\author{
\authorblockN{Neri Merhav}
\authorblockA{
  Department of Electrical Engineering\\
    Technion -- Israel Institute of Technonolgy\\
    Technion City, Haifa 32000, Israel\\
     Email: merhav@ee.technion.ac.il
\vspace{-3ex}
}}

\maketitle

\vspace{-.5in}

\begin{abstract}
We provide a simple physical interpretation, in the context of the second law of
thermodynamics, to the 
information inequality (a.k.a.\ the Gibbs' inequality, which is also
equivalent to the log--sum
inequality), asserting that the relative entropy between two probability distributions
cannot be negative. Since this inequality stands at the basis of the data
processing theorem (DPT), and the DPT in turn is at the heart of most, if not
all, proofs of converse theorems in Shannon theory, it is observed that
conceptually, the
roots of fundamental limits of Information Theory can actually be attributed
to the laws of physics, in particular, the second law
of thermodynamics, and indirectely, also the law of energy conservation.
By the same token, in the other direction: one can view the second law as
stemming from
information--theoretic principles.
\end{abstract}


\section{Introduction}

While the laws of physics draw the boundaries between the possible and the
impossible in Nature, the coding theorems of Information Theory, or more
precisely, their converse parts, draw the boundaries between the possible
and the impossible in the design and performance of coded communication
systems and in data processing.
A natural question that may arise, in view of these two facts, is whether there
is any relationship between them. It is the purpose of this work to touch upon
this question and to make an attempt to provide at least a partial answer.

Perhaps the most fundamental inequality in Information Theory is the
so called {\it information inequality} (cf.\ e.g., \cite[Theorem 2.6.3, p.\
28]{CT06}), which asserts that the relative entropy (a.k.a.\ the
Kullback--Leibler divergence) between two probability distributions over the
same alphabet $P=\{P(x),~x\in\calX\}$ and
$Q=\{Q(x),~x\in\calX\}$,
$$D(P\|Q)=\sum_{x\in\calX}P(x)\log\frac{P(x)}{Q(x)},$$
can never be negative, and a similar fact applies to probability density
functions with the summation across $\calX$ being replaced by integration.

The {\it log--sum inequality} (LSI)
\cite[Theorem 2.7.1, p.\ 31]{CT06},
which asserts that
for two sets of non--negative numbers,
$(a_1,a_2,\ldots,a_n)$ and  
$(b_1,b_2,\ldots,b_n)$:
$$\sum_{i=1}^n a_i\log\frac{a_i}{b_i}\ge \left(\sum_{i=1}^n
a_i\right)\log\left(\frac{\sum_{i=1}^n a_i}{\sum_{i=1}^n b_i}\right),$$
is completely equivalent\footnote{The information inequality is obtained
from the LSI when $(a_1,a_2,\ldots,a_n)$ and $(b_1,b_2,\ldots,b_n)$ both sum
to unity, and conversely, the LSI is obtained from the information
inequality, by applying the latter to the probability distributions $P_i=a_i/\sum_ja_j$
and $Q_i=b_i/\sum_jb_j$.}
to the information inequality, although proved in
\cite{CT06} in a rather different manner. 

Yet another name for the same
inequality, which is more
frequently encountered in the jargon of physicists, is the 
{\it Gibbs' inequality}: 
When the information inequality is 
applied to two probability distributions of the Boltzmann form (cf.\
Section IV below), it yields an interesting inequality 
concerning their corresponding free energies (cf.\ e.g., 
\cite[Section 5.6, pp. 143--146]{Kardar07}), which serves as
a useful tool for obtaining good bounds on the free energy of a complex system, when 
its exact value is difficult to calculate.

In this work, we provide a simple physical interpretation to this inequality
of the the free energies, and thereby also to the information inequality, or
the log--sum inequality. This physical interpretation is directly related to
the second law of thermodynamics, which asserts
that the entropy of an isolated physical 
system cannot decrease: According to this interpretation, the divergence between two
probability distributions is proportional to the energy dissipated in the
system when it
undergoes an irreversible process, and hence converts this energy loss into
entropy production,
or heat. Thus, the non--negativity of the relative entropy is
related to the non--negativity of this entropy change, which is, as said, the
second law of thermodynamics.

Since the mutual information can be thought of as an instance of the relative
entropy, and so can the difference between two mutual informations defined along a
Markov chain, then the data processing theorem (DPT) can, of course, also be given the
very same physical interpretation. Considering the fact that the DPT is
pivotal to most, if not all, converse theorems in Information Theory, this
means that, in fact, the fundamental limits of Information Theory can, at
least conceptually, be attributed to the laws
of physics,
in particular, to the second law of 
thermodynamics:\footnote{Another law of physics
that plays a role here, at least indirectly, is the law of energy
conservation, because our derivations are all based on the Boltzmann--Gibbs
distribution of equilibrium statistical mechanics, and this distribution, in
turn, is derived on the basis of the energy conservation law.}
The rate loss in any suboptimal coded communication system,
is given the meaning of irreversibility and entropy production in a
corresponding physical system. Optimum
(or nearly optimum) communication systems are corresponding to reversible
processes (or lack of any process at all) with no entropy production. Stated
in somewhat different
words, had there been a communication system that violated a fundamental
limit (e.g., beating the entropy, or channel capacity), then in principle, one could
have constructed a physical system that
violates the second law, and vice versa.

The outline of the remaining part of the paper is as follows.
In Section II, we give some basic back background in statistical physics.
Section III reviews the role of the DPT in many of the converse theorems
in the Shannon theory. In Section III, we offer a physical interpretation
to the Gibbs' inequality and show how it applies to the DPT in two different
scenarios. Finally, in Section IV, we discuss relationships between reversible
processes in physics and error exponents of classical Neyman--Pearson hypothesis testing.

\section{Physics Background}
\label{s:physbg}

Consider a physical system with $n$ particles, which at any time instant, can
be found in any one out of a
variety of microscopic states (or {\it micorstates}, for short). 
The microstate is defined by the full physical information 
about all $n$ particles, e.g., the positions, momenta, angular momenta,
spins, etc., depending on the type of
the physical system. In particular, a microstate is designated by
$\bx=(x_1,x_2,\ldots,x_n)$,
where each $x_i$ may itself be a vector, consisting of all the relevant
physical state variables (such as the above) for particle number $i$ at a given time
instant. Associated with every microstate $\bx$,
there is an energy function, a.k.a.\ the
{\it Hamiltonian}, $\calE(\bx)$. 
For example, in the case of the ideal gas, $x_i=(\bp_i,\br_i)$, where $\bp_i$
and $\br_i$, both three dimensional vectors, are the momentum and the position
of particle number $i$, respectively, and
\begin{equation}
\label{idealgas}
\calE(\bx)=\sum_{i=1}^n\left[\frac{\|p_i\|^2}{2m}+mgz_i\right],
\end{equation}
where $m$ is the mass of
each particle, $g$ is the gravitation constant, and $z_i$ is the height -- one
of the components of $\br_i$. 

One of
the most fundamental results in statistical physics
(based on the law of energy conservation and the postulate
that all microstates of the same energy are equiprobable)
asserts that, when a system lies in thermal equilibrium with the environment (heat
bath), the probability of finding the system at state
$\bx$ is given by the {\it
Boltzmann--Gibbs} distribution
\begin{equation}
\label{bd}
P(\bx)=\frac{e^{-\beta\calE(\bx)}}{Z(\beta)}
\end{equation}
where $\beta=1/(kT)$, $k$ being Boltzmann's constant 
and $T$ being temperature,
and $Z(\beta)$ is the normalization constant,
called the {\it partition function}, which
is given by
\begin{equation}  \label{eq:Z}
Z(\beta)=\sum_{\bx} e^{-\beta\calE(\bx)}, \text{ or }
Z(\beta)=\int \mbox{d}\bx\, e^{-\beta\calE(\bx)},  
\end{equation}
depending on whether $\bx$ is discrete or continuous.
The partition function is
a key quantity from which many
important macroscopic physical quantities can be derived.
For example, the average internal energy w.r.t. (\ref{bd}) is
\begin{equation}
\label{energy}
E=\bE\{\calE(\bX)\}=-\frac{\mbox{d}\ln Z(\beta)}{\mbox{d}\beta},
\end{equation}
the entropy (in units of $k$) pertaining to 
\eqref{bd} is 
\begin{equation}
\label{entropy}
\Sigma(\beta)\eqd \frac{S(\beta)}{k}=-\bE\{\ln P(\bX)\}=\ln Z(\beta)+\beta\cdot E,
\end{equation}
and the {\it free energy} is given by
\begin{equation}
\label{freeenergy}
F(\beta)=-\frac{\ln Z(\beta)}{\beta}.
\end{equation}
From eq.\ (\ref{entropy}), one readily obtains the well known relationship
$$F=E-ST.$$ 
Thus, any change in the internal energy, along
a fixed temperature (isothermal) process, 
is given by 
$$\Delta E=\Delta F+T\Delta S,$$ 
in other words, it 
consists of two components: the first is the 
change in the free energy, $\Delta F$, and 
the second pertains to entropy production,
$T\Delta S$. By the first law of thermodynamics, which is
actually, the law of energy conservation, 
$$\Delta E=\Delta Q+\Delta W,$$
namely, the origins of any change in the internal energy may be a combination
of the heat $\Delta Q$
transferred into the system
and the work $\Delta W$ applied to it. 
According to the thermodynamical definition, 
the entropy difference, $\Delta S$,  between 
two macroscopic states $A$ and $B$, is defined as 
$\int_A^B\mbox{d}Q/T$, where the integration is along
a {\it quasi--static} or {\it reversible} process, i.e., a process that is
slow enough such that, along the way, the system is kept always very close to
equilibrium. By
the {\it Clausius theorem} (cf.\ e.g., \cite[p.\ 13]{Kardar07}),
in the above described 
isothermal process, $\Delta S$ is never smaller than $\Delta Q/T$,
with equality when the process is reversible.
Thus, by comparing the two expressions of $\Delta E$, we
immediately observe that $\Delta W\ge\Delta F$.

The free energy is then given a meaning 
of crucial importance in thermodynamics and
statistical physics: The difference, $\Delta F$, between the free energies
associated with
two equilibirium points pertaining to the same temperature (but with two
different values
of some other control parameter, such as pressure or magnetic field) has the
physical meaning of the {\it minimum} amount of work that should be applied to
the system in order to transfer it
between these two equilibria
along an isothermal process, and this
minimum is attained when the process is
reversible.\footnote{This fact is also known as the {\it minimum work
principle}.}
Equivalently, the negative free--energy difference, $-\Delta F$, is the
{\it maximum} amount of work that can be exploited from the system in an
isothermal process, and this maximum is achieved, again, if the process is reversible.
The second law of thermodynamics, as mentioned earlier, asserts that the entropy of an isolated
system cannot decrease.

\section{The Data Processing Theorem and Fundamental Limits}
\label{s:th}

As mentioned earlier, our observations apply to any fundamental limit, or
converse theorem, that makes use of the information inequality, in one way or another.
However, even if we confine our attention only to those that use it explicitely 
in the form of the DPT,
it is not difficult to appreciate the fact that we already cover many of the
fundamental limits, if not all of them. Here are just a few examples.\\

\noindent
{\it Lossy/lossless source coding}: Consider a source vector $U^N=(U_1,\ldots U_N)$
compressed into a bitstream $X^n=(X_1,\ldots,X_n)$ from which the decoder
generates a reproduction $V^N=(V_1,\ldots,V_N)$ with distortion
$\sum_{i=1}^N \bE\{d(U_i,V_i)\}\le ND$. Then, by the DPT,
$I(U^N;V^N)\le I(X^n;V^N)\le H(X^n)$, 
where $I(U^N;V^N)$ is further lower bounded by
$NR(D)$ and $H(X^n)\le n$, which together lead to the converse to the lossy data
compression theorem, asserting that the compression ratio $n/N$ cannot be
less than $R(D)$. Lossless compression is obtained, of course, as a special
case where $D=0$.\\

\noindent
{\it Channel coding under bit error probability}: Let $U^N=(U_1,\ldots U_N)$
be drawn from the binary symmetric course (BSS), designating $M=2^N$ equiprobable
messages of length $N$. The encoder maps $U^N$ into a channel input vector
$X^n$, which in turn, is sent across the channel. The receiver observes $Y^n$,
a noisy version of $X^n$, and decodes the message as $V^N$. Let
$P_b=\frac{1}{N}\sum_{i=1}^N\mbox{Pr}\{V_i\ne U_i\}$ designate 
the bit error probability. Then, by the DPT, $I(U^N;V^N)\le I(X^n;Y^n)$, where
$I(X^n;Y^n)$ is further upper bounded by $nC$, $C$ being the channel capacity,
and $I(U^N;V^N)=H(U^N)-H(U^N|V^N)\ge N-\sum_{i=1}^N H(U_i|V_i)\ge
N-\sum_ih_2(\mbox{Pr}\{V_i\ne U_i\})\ge N[1-h_2(P_b)]$.
Thus, for $P_b$ to vanish, the coding rate, $N/n$ should not exceed $C$.\\

\noindent
{\it Channel coding under block error probability -- Fano's inequality}: 
This is the same as in the previous item, except that the error performance is the block
error probability $P_B=\mbox{Pr}\{V^N\ne U^N\}$. This time, $H(U^N|V^N)$,
which is identical to $H(U^N,E|V^N)$, with $E\eqd \calI\{V^N\ne U^N\}$
($\calI$ being the indicator function), is
decomposed as $H(E|V^N)+H(U^N|V^N,E)$, where the first term is upper bounded
by 1 and the second term is upper bounded by $P_B\log(2^N-1) < NP_B$, owing to
the fact that the maximum of $H(U^N|V^N,E=1)$ is obtained when $U^N$ is
distributed uniformly over all $V^N\ne U^N$. Putting these facts all together,
we obtain Fano's inequality $P_B\ge 1-1/n-C/R$, where $R=N/n$ is the coding
rate. Thus, the DPT directly supports Fano's inequality,
which in turn is the main tool for proving converses to channel
coding theorems in a large variety of communication situations, including
network configurations.\\

\noindent
{\it Joint source--channel coding and the separation principle}: In a joint
source--channel situation, where the source vector $U^N$ is mapped into a
channel input vector $X^n$ and the channel output vector $Y^n$ is decoded into
a reconsdtruction $V^N$, the DPT gives rise to the chain
of inequalities $NR(D)\le I(U^N;V^N)\le I(X^n;Y^n)\le nC$, which is the
converse to the joint source--channel coding theorem, whose direct part can
be achieved by separate source- and channel coding. The first two examples above are special
cases of this.\\

\noindent
{\it Conditioning reduces entropy}: Perhaps even more often than the
term ``data processing theorem'' can be found as part of a proof of
a converse theorem,  one encounters an equivalent of this theorem 
under the slogan ``conditioning reduces entropy''. This in turn is part of virtually every 
converse proof in the literature. Indeed, if $(X,U,V)$ is a triple of RV's,
then this statement means that $H(X|V)\ge H(X|U,V)$. If, in addition, $X\to U\to V$
is a Markov chain, then $H(X|U,V)=H(X|U)$, and so, $H(X|V)\ge H(X|U)$, which
in turn is equivalent to the more customary form of the DPT, $I(X;U)\ge
I(X;V)$, obtained by subtracting $H(X)$ from both sides of the entropy
inequality. In fact, as we shall see shortly, it is this entropy inequality
that lends itself more naturally to a
physical interpretation. Moreover, we can think of the
conditioning--reduces--entropy inequality as another form of the DPT even in
the absence of the aforementioned Markov condition, because $X\to(U,V)\to V$ is always a
Markov chain.

\section{Physics of the Information Inequality \& DPT}

We consider two forms of the information inequality an the DPT,
one corresponding to an isothermal
process and one -- to an adiabatic process (fixed
amount of heat).

\subsection{Isothermal Version}

Consider a system, with a microstate $\bx$,
which may have two possibile Hamiltonians -- $\calE_0(\bx)$ and
$\calE_1(\bx)$.
Let $Z_i(\beta)$, denote the partition 
function pertaining to $\calE_i(\cdot)$,
that is,
$Z_i(\beta)=\sum_{\bx} e^{-\beta\calE_i(\bx)}$, $i=0,1$,
where $\beta=1/(kT)$ is the inverse temperature.
Since $\beta$ is fixed throughout this section, we will
also use the shorthand notation $Z_i$ for the partition function.
Let $P_i(\bx)$ denote the Boltzmann--Gibbs distribution 
(cf.\ eq.\ (\ref{bd})) pertaining to $Z_i$, $i=0,1$ (both for the same
given value of $\beta$).
Applying the information inequality to $P_0$ and $P_1$, we get:
\begin{eqnarray}
0&\le&D(P_0\|P_1)=\sum_{\bx}P_0(\bx)\ln\left[\frac{e^{-\beta\calE_0(\bx)}/Z_0}
{e^{-\beta\calE_1(\bx)}/Z_1}\right]\nonumber\\
&=&\ln Z_1-\ln
Z_0+\beta\bE_0\{\calE_1(\bX)-\calE_0(\bX)\}
\end{eqnarray}
where $\bE_0\{\cdot\}$ denotes the expectation operator w.r.t.\ $P_0$.
After a minor algebraic rearrangement, this becomes:
\begin{eqnarray}
\label{Wgedeltaf}
\bE_0\{\calE_1(\bX)-\calE_0(\bX)\}&\ge& kT\ln Z_0-kT\ln Z_1\nonumber\\
&\equiv& F_1-F_0,
\end{eqnarray}
where $F_i$ is the free energy pertaining to
$P_i$, $i=0,1$ (cf.\ eq.\ \ref{freeenergy})).

We now offer the following physical interpretation to this inequality:
Imagine that a
system with Hamiltoinan $\calE_0(\bx)$ is in equilibrium
for all $t < 0$,\footnote{Since
the information inequality applies to any pair of distributions, it is
conceivable that the interpretation we offer may remain relevant even beyond
the realm of systems in equilibirium. Indeed, even if the system is
away from equilibrium, when it is
nevertheless in steady state (in the sense that
macroscopic physical quantities are time--invariant), the negative logarithm
of the density function can be given the meaning of an {\it effective
Hamiltonian} \cite{KNST09}.
This, however, is beyond the scope of this
work.}
but then, at time $t=0$, the Hamitonian changes {\it abruptly} from the
$\calE_0(\bx)$ to $\calE_1(\bx)$ (e.g., by 
suddenly applying a force, like pressure or a magnetic field, to the
system), which means that if the system is found at state
$\bx$ at time $t=0$, additional energy of $W(\bx)=\calE_1(\bx)-\calE_0(\bx)$ is
suddenly `injected' into it. This additional energy can be thought of as work
performed on the system, or as supplementary potential energy.
Of course, $W(\bx)$ is a random variable due to the randomness of $\bx$.
Since this passage between $\calE_0$ and $\calE_1$ is abrupt,
and the microstate $\bx$ does not change instantaneously,
the expectation of $W(\bX)$ should be taken
w.r.t.\ $P_0$, and 
this average is exactly what we have at the left--hand side eq.\
(\ref{Wgedeltaf}).
The Gibbs' inequality tells us then that
this average work is at least as large as
$\Delta F=F_1-F_0$, the increase in free energy, 
in compliance to the explanation in Section II.
The difference 
$$\bE_0\{W(\bX)\}-\Delta F=kT\cdot D(P_0\|P_1)\ge 0$$ 
is due to the irreversible nature
of this abrupt energy injection, and this irreversibility
means an increase of the total entropy 
of the system and its environment.\footnote{See also \cite{PKL08}, \cite{KPV07}, \cite{HJ09}
and references therein, where the same conclusions are reached
from a more general perspective of irrreversible processes, but under
certain limiting assumptions on the physical system.}
Thus, the Gibbs'
inequality is, in fact, a version of the second law of
thermodynamics, and the relative entropy is given a very simple physical
significance. We next consider two examples.\\

\noindent
{\it Example 1 -- Fixed--to-variable compression and the Ising model.}
A natural information--theoretic example for this can be 
easily motivated by the interpretation of
the relative entropy as the rate loss (or, the redundancy) due to mismatch in 
fixed--to--variable lossless data compression: Suppose that $\bX\in\{-1,+1\}^n$ emerges from
a first--order Markov source $P_0(\bx)=\prod_{i=1}^nP_0(x_i|x_{i-1})$,
where 
$$P_0(x|x')= \frac{\exp\{Jx\cdot x'\}}{Z_0},~~
x,x'\in\{-1,+1\},$$
and where $J$ is
a given constant and
$$Z_0=2\cosh(J).$$
However, the code designer designs a 
Shannon code according to $P_1(\bx)=\prod_{i=1}^n P_1(x_i|x_{i-1})$, where
$$P_1(x|x')= \frac{\exp\{Jx\cdot x'+Kx\}}{\zeta(x')},~~x,x'\in\{-1,+1\}$$ 
where $K$ is another
given constant and $\zeta(x)$ is the appropriate normalization factor
given by
$$\zeta(x)=\left\{\begin{array}{ll}
2\cosh(J+K) & x=+1\\
2\cosh(J-K) & x=-1\end{array}\right.$$
Considering the fact that $x\in\{-1,+1\}$, $\zeta(x)$ can also
be written in a unified way as
$$\zeta(x)=Z_1\cdot
\left[\frac{\cosh(J+K)}{\cosh(J-K)}\right]^{x/2}.$$
where
$$Z_1=2\sqrt{\cosh(J+K)\cosh(J-K)}.$$
From the physics point of view, both $P_0$ and $P_1$ can be thought of as
Boltzmann--Gibbs distributions with inverse temperature $\beta=1$: For the
former, we define the Hamiltonian as
\begin{eqnarray}
\calE_0(\bx)&\eqd&-n\ln Z_0-
\sum_{i=1}^n\ln P_0(x_i|x_{i-1})\nonumber\\
&=&-J\cdot\sum_ix_{i-1}x_i
\end{eqnarray}
which can be
thought of as the energy pertaining to nearest--neighbor interactions 
between spins in a one--dimensional array, that is,
the one--dimensional {\it Ising model} (see, e.g., 
\cite[Sect.\ 1.8]{Baxter82})
with a coupling coefficient $J$, in the absence of a
magnetic field.
On the other hand,
for $P_1$ we define:
\begin{eqnarray}
\calE_1(\bx)&\eqd&-n\ln Z_1
-\sum_{i=1}^n\ln P_1(x_i|x_{i-1})\nonumber\\
&=&-J\sum_ix_{i-1}x_i-K\sum_ix_i-\nonumber\\
& &\frac{1}{2}\left[\ln\frac{\cosh(J-K)}{\cosh(J+K)}\right]
\cdot\sum_ix_{i-1}\nonumber\\
&\approx&-J\sum_ix_{i-1}x_i-\nonumber\\
& &\left(K+
\frac{1}{2}\ln\frac{\cosh(J-K)}{\cosh(J+K)}\right)
\cdot\sum_ix_i\nonumber\\
&\eqd&-J\sum_ix_{i-1}x_i-B\sum_i x_i
\end{eqnarray}
where in the approximate equality we neglected ``edge effects''
that make the (relatively)
small difference between $\sum_ix_i$ and $\sum_ix_{i-1}$
(for large $n$). This is
the same Ising model as before, but now also with a magnetic field $B$.
Thus, 
$$\calE_1(\bx)-\calE_0(\bx)=-B\sum_ix_i$$ 
is the
energy injected by an abrupt application of the magnetic field $B$. We
have therefore demonstrated that the
entropy production due to the irreversiblilty of this abrupt magnetic field
is (within the additive constant, $\Delta F=1\cdot(\ln Z_0-\ln Z_1)$)
proportional to the redundancy of the mismatched code.\\

\noindent
{\it Example 2 -- Run--length coding and the grand--canonical ensemble.}
The Boltzmann--Gibbs distribution of eq.\ (\ref{bd}),
a.k.a.\ the {\it canonical distribution}, is the equilibrium
distribution of a system that is allowed to exchange heat energy with its
environment at a fixed temperature $T$. It also assumes that the system
has a fixed number of particles $n$, and a fixed volume $V$,
whenever the volume is a relevant factor. 

When the system is allowed
to exchange with the environment, 
not only energy, but also matter, namely, particles, then
eq.\ (\ref{bd}) is extended to the {\it grand--canonical distribution}
\cite[Sect.\ 4.9]{Kardar07},
whose microstate is defined as $(\bx,n)$, where $n$ is now a random variable, and
$\bx$ is defined as before for the given $n$. According to this
distribution,
$$P(\bx,n)=\frac{e^{\beta(\mu n-\calE(\bx))}}{\Xi(\beta,\mu)}$$
where 
$$\Xi(\beta,\mu)=\sum_{n\ge 0} e^{\beta\mu n}\sum_{\bx} e^{-\beta\calE(\bx)}
\eqd \sum_{n\ge 0} e^{\beta\mu n}Z(\beta,n)$$
is the {\it grand partition function}. 
The parameter $\mu$, which
is called the {\it chemical potential},
controls the average number of particles in the system.
Note that $P(\bx,n)$ can be
thought of as $P(n)\cdot P(\bx|n)$ where $P(\bx|n)$ obeys the
canonical distribution for the given $n$ 
and $P(n)$ is proportional to $e^{\beta\mu
n}Z(\beta,n)$. It is well known (see, e.g., \cite{Kardar07})
that $kT\ln\Xi(\beta,\mu)$ gives the equilibrium pressure--volume product of
the system, $\bP V$. Now let $P_0(\bx,n)$ and $P_1(\bx,n)$ be two
grand--canonical distributions that differ only in the chemical potentials,
$\mu_i$, $i=0,1$, respectively. 
Applying the information inequality, we get
\begin{eqnarray}
0&\le&D(P_0\|P_1)\nonumber\\
&=&\ln \Xi(\beta,\mu_1)-\ln \Xi(\beta,\mu_0)+\nonumber\\
& &\beta(\mu_0-\mu_1)\bE_0\{N\}
\end{eqnarray}
where $N$ designates the random number of particles.
Dividing by $\beta$ and rearranging terms, this becomes:
$$\bP_1V \ge \bP_0V+(\mu_1-\mu_0)\bE_0\{N\},$$
and after dividing by $V$ (which is assumed fixed), we get:
$$\bP_1 \ge \bP_0+(\mu_1-\mu_0)\bE_0\{\rho\},$$
where $\rho=N/V$ is the density of particles.

A natural information--theoretic analogue of this is run--length coding:
Given a $0$--$1$ binary memoryless source with a very high probability of `0',
which we shall designate by $e^{\mu}$ ($\mu < 0$, $\beta=1$), the idea is to
encode the number $N$ of successive zeroes between every two 
consecutive ones.
Clearly, the distribution of $N$ is exponential
$$\mbox{Pr}\{N=n\}=\frac{e^{\mu n}}{\Xi(\mu)}$$
where, with a slight abuse of notation, we define
$$\Xi(\mu)=\frac{1}{1-e^{\mu}},$$
and where we have assumed $\calE(\bx)=-\ln P(\bx|n)$, and so, $Z(1,n)=1$ for all
$n$. Thus, when applying run--length coding, the price of mismatch in $\mu$
is parallel to the difference between the two sides of the above
pressure inequality, where the `pressure' in run--length coding
is proportional to $-\ln (1-e^{\mu})$. As $\mu\uparrow 0$, the
pressure increases, and more `particles' (i.e., runs of zeroes) enter into
the system, which means that the runlengths becomes larger.
Thus, we have demonstrated an analogy between run--length coding and
the physics of the grand--canonical ensemble: the
log--probability of `0' plays the role the chemical potential whereas the
log--probability of `1' is associated with pressure.
This concludes Example 2.\\

Returning to the general framework,
let us now see how the Gibbs' inequality is related to the DPT. 
Consider a triple of random variables $(X,U,V)$ which form a Markov chain $X\to U\to V$. 
The DPT asserts that $I(X;U)\ge I(X;V)$. 
We can obtain the DPT as a special case of the Gibbs' inequality because
\begin{eqnarray}
I(X;U)-I(X;V)&=&H(X|V)-H(X|U)\nonumber\\
&=&\bE\{D(P_{X|U,V}(\cdot|U,V)\|P_{X|V}(\cdot|V))\}\nonumber
\end{eqnarray}
where the expectation is w.r.t.\ the randomness of $(U,V)$. Thus,
For a given realization $(u,v)$ of $(U,V)$, consider
the Hamiltonians
$\calE_0(x)=-\ln P(x|u)=-\ln P(x|u,v)$
and
$\calE_1(x)=-\ln P(x|v)$,
pertaining to a single `particle' whose state is $x$.
Let us also set $\beta=1$. Thus, for a given $(u,v)$:
\begin{eqnarray}
\bE_0\{W(X)\}&=&\sum_x P(x|u,v)[\ln P(x|u)-\ln P(x|v)]\nonumber\\
&=&H(X|V=v)-H(X|U=u)
\end{eqnarray}
and after further averaging w.r.t.\ $(U,V)$, the average work becomes $H(X|V)-H(X|U)=I(X;U)-I(X;V)$. 
Concerning the free energies, we have
\begin{eqnarray}
Z_0(1)&=&\sum_{x} \exp\{-1\cdot[-\ln P(x|u,v)]\}\nonumber\\
&=&\sum_{x} P(x|u,v)=1
\end{eqnarray}
and similarly,
$$Z_1(1)=\sum_{x} P(x|v)=1$$
which means that $F_0(1)=F_1(1)=0$, and so $\Delta F=0$ as well. 
So by the Gibbs' inequality, the
average work, $I(X;U)-I(X;V)$, cannot be 
smaller than the free--energy difference, which
in this case vanishes, namely, $I(X;U)-I(X;V)\ge 0$, which is the DPT.
Note that in this case, there is a maximum degree of irreversibility: 
The identity $I(X;U)-I(X;V)=H(X|V)-H(X|U)$ means that
whole average work, $W=I(X;U)-I(X;V)$, goes for entropy 
increase $T\Delta \Sigma=1\cdot[H(X|V)
-H(X|U)]$, whereas the free energy remains unchanged, as mentioned
earlier. Moreover, the entire entropy increase goes to the system under
discussion, and none
of it goes to the environment. 

At this point a comment is in order:
The rate loss of a suboptimal communication
system, when viewed from the DPT perspective,
may be attributed to two possible factors: one factor 
comes from a possible mismatch between
actual distributions and optimum distributions in the
information--theoretic sense, for example,
the encoder may not induce the capacity--achieving
channel input distribution or the test channel of the rate--distortion
function. The other factor is a possible gap between mutual
informations along the Markov chain 
($I(X;U)$ may be strictly larger than $I(X;V)$), which
actually means {\it information loss}, and which is
{\it irreversible} ($U$ cannot be retreived from $V$). 
It is the latter kind of loss that is parallel to the irreversible
free energy loss and dissipation. 

From a more general physical perspective, 
we can think of the Hamiltonian
$$\calE_\lambda(\bx)=\calE_0(\bx)+\lambda[\calE_1(\bx)-\calE_0(\bx)]$$
as a linear interpolation between the two extremes, $\lambda=0$ and
$\lambda=1$, pertaining to $\calE_0$ and $\calE_1$, 
and then $\lambda$ can be thought of as a control parameter
or a `force' that influences the system.
The Jarzynsky equality 
(cf.\ e.g., \cite{PKL08} and references therein) tells
that under certain conditions on the system and the environment,
and given any protocol for a temporal change in $\lambda$, 
designated by $\{\lambda_t\}$,
for which $\lambda_t=0$ for all $t<0$, and
$\lambda_t=1$ for all $t\ge \tau$ ($\tau \ge 0$),
the work $W$ applied to the system
is a RV that satisfies 
$$\bE\{e^{-\beta W}\} = e^{-\beta\Delta F}.$$
By Jensen's inequality, 
$$\bE\{e^{-\beta W}\}\ge
\exp(-\beta\bE\{W\}),$$ which then gives $\bE\{W\}\ge\Delta F$,
for an arbitrary protocol $\{\lambda_t\}$. 
The Gibbs' inequality is then a special
case, where $\lambda_t$ is given by the unit step function, but
it applies regardless of 
the assumptions of \cite{PKL08}. At the other extreme, when
$\lambda_t$ changes very slowly, corresponding to a reversible process, $W$
approaches determinism, and then Jensen's inequality becomes tight. In the
limit of an arbitrarily slow process, 
this yields $W=\Delta F$, with no increase in entropy.

\subsection{Adiabatic Version}

Thus far, we discussed an isothermal process,
where the change was attributed to the
Hamiltonian  -- a transition from $\calE_0$ to $\calE_1$.
In the special case where the two Hamiltonians are proportional to
one another, namely, when 
$\calE_1(x)/\calE_0(x)=\mbox{const.}$, independent of $x$,
one can, of course, still consider it as an 
isothermal process and refer the change in the Hamiltonian to
that of a multiplicative control parameter 
$\lambda$, as before (e.g., the 
harmonic potential $\frac{\lambda}{2}x^2$). 
But perhaps even more natural, in this case,
is to refer the change to 
temperature. In this case, there is no 
external mechanical work, and the change in
the internal energy of the system comes solely from heat: We replace a
heat bath (large environement) with temperature $T_0=1/(k\beta_0)$ by a
heat bath with a higher temperature
$T_1=1/(k\beta_1)$. If we apply the Gibbs' inequality to this special case,
this amounts to
$$\ln Z(\beta_1)\ge \ln
Z(\beta_0)+(\beta_0-\beta_1)\bE_0\{\calE_0(X)\}$$
which is easily shown (cf.\ eq.\ (\ref{entropy})) to be equivalent to
\begin{eqnarray}
\Delta \Sigma&\equiv&\Sigma(\beta_1)-\Sigma(\beta_0)\nonumber\\
&\ge&
\beta_1[\bE_1\{\calE_0(X)\}-\bE_0\{\calE_0(X)\}]\equiv
\frac{\Delta Q}{kT_1},
\end{eqnarray}
where $\Sigma(\beta_0)$ and $\Sigma(\beta_1)$ are the 
equilibrium entropies (in units of $k$) pertaining to
$\beta_0$ and $\beta_1$, respectively, and $\Delta Q$ is the amount of
heat injected into the system, assuming there is no mechanical work.
This inequality is a special case of the Clausius theorem (mentioned
earlier), which in its general form,
asserts that $\Delta S=k\Delta\Sigma$ is never smaller
than $\int dQ/T$ for any process, with equality in the case of a 
reversible process. The expression $\Delta Q/T_1$
is the result of this integral when the heat bath of temperature $T_0$
is abruptly replaced by one with temperature $T_1$. An alternative
interpretation of this inequality is, again, as
an instance of the second law: The entropy of our
system increases by $\Delta S$ and the entropy of the (new) heat bath
decreases by $\Delta Q/T_1$, thus the net entropy change of the
combined system (which is assumed isolated), 
$\Delta S-\Delta Q/T_1$, must be non--negative.

In the information--theoretic context, the relevant
situation is one where $P(x|u,v)=P(x|u)$ 
and $P(x|v)=\int\mbox{d}uP(x|u,v)P(u|v)$ can be represented as Boltzmann
distributions with the same Hamiltonian, but which may differ in temperature and
possibly in shifts (by $u$ or $v$). I.e.,
$$P(x|u,v)=P(x|u)=\frac{e^{-\beta_0\calE(x-u)}}{Z(\beta_0)};$$
$$P(x|v)=\frac{e^{-\beta_1\calE(x-v)}}{Z(\beta_1)}~~~\beta_1 < \beta_0$$
This turns out to be the case when $X$,
$U$ and $V$ are related by a cascade of two additive channels of the
same family (e.g., a degraded broadcast channel), 
one from $V$ to $U$ and the other from $U$ to $X$ (or in the other
direction). Two classical examples are
those when both channels are binary and symmetric (with possibly two different
crossover parameters), and when they are both Gaussian (with
possibly different noise variances). Other examples of these properties
could pertain to any choice of an infinitely divisible random variable as
a noise model in both channels, like 
the Poisson RV, the binomial RV, and so on.

Using again the Gibbs' inequality as before, 
we now get, for given $u$ and $v$:
\begin{eqnarray}
\ln Z(\beta_1)
&\ge&\ln
Z(\beta_0)+\beta_0\bE_{\beta_0,u,v}\{\calE(X-u)\}-\nonumber\\
& &\beta_1\bE_{\beta_0,u,v}\calE(X-v),
\end{eqnarray}
where $\bE_{\beta_0,u,v}$ 
denotes expectation w.r.t.\ $P(x|u,v)$ as defined above.
Now, assuming shift--invariance of integrals 
over $x$ (as is the case in the BSC and Gaussian examples mentioned above), 
$\bE_{\beta_0,u,v}\{\calE(X-u)\}=
\bE_{\beta_0,0,0}\{\calE(X)\}
\eqd\bE_{\beta_0}\{\calE(X)\}$, independently
of $u$ and $v$. 
As for the third term, from the above relation between $P(x|u,v)$ and 
$P(x|v)$, it is apparent that
after averaging $\bE_{\beta_0,u,v}\{\calE(X-v)\}$ (which is independent of $u$) w.r.t.\ $P(u|v)$, it
becomes $\bE_{\beta_1,v}\{\calE(X-v)\}=
\bE_{\beta_1,0}\{\calE(X)\}\eqd
\bE_{\beta_1}\{\calE(X)\}$. Thus, we get
\begin{eqnarray}
\Sigma(\beta_1)&\equiv&\ln Z(\beta_1)+\beta_1\bE_{\beta_1}\calE(X)\}\nonumber\\
&\ge&\ln Z(\beta_0)+\beta_0\bE_{\beta_0}\{\calE(X)\}\equiv \Sigma(\beta_0)
\end{eqnarray}
This is then a special case of the inequality
$\Delta \Sigma\ge \Delta Q/(kT_1)$, where $\Delta Q=0$, namely,
an {\it adiabatic process}, and then $\Delta \Sigma \ge 0$, or $\Delta S\ge 0$.
The information loss due to the DPT
again has the physical interpretation of entropy 
increase, but this time it
is purely due to temperature increase,
rather than the dissipated work
that we have seen before.

We end this section with two simple examples,
namely, the Gaussian broadcast channel and the binary symmetric broadcast
channel. In both examples, 
we view the mutual information difference, which is the entropy increase,
as an integral of temperature, and thereby identify the corresponding
heat capacity from the integrand.\\

\noindent
{\it Example 3 -- Gaussian degraded broadcast channel:}
Consider a Gaussian degraded broadcast channel, i.e.,
a cascade of two independent additive white Gaussian noise (AWGN) channels,
given by:
$$X=U+\frac{N_1}{\sqrt{\beta_0}}$$ 
and 
$$U=V+N_2\sqrt{\frac{1}{\beta_1}-\frac{1}{\beta_0}},~~~~~
\beta_1 <\beta_0,$$
where $N_1$ and
$N_2$ are both zero--mean, unit--variance Gaussian RV's, 
independent of each other as well as of $V$, 
which in turn has an arbitrary density with
$\bE\{V^2\}< \infty$. In this case,
\begin{eqnarray}
\Delta \Sigma&=&I(X;U)-I(X;V)\nonumber\\
&=&h(X|V)-h(X|U)\nonumber\\
&=&\frac{1}{2}\ln\left(\frac{2\pi e}{\beta_1}\right)-
\frac{1}{2}\ln\left(\frac{2\pi e}{\beta_0}\right)\nonumber\\
&=&\frac{1}{2}\ln\frac{\beta_0}{\beta_1}\nonumber\\
&=&\frac{1}{2}\int_{\beta_1}^{\beta_0}\frac{\mbox{d}\beta}{\beta}\nonumber\\
&=&\int_{T_0}^{T_1}\frac{\mbox{d}T}{2T},\nonumber
\end{eqnarray}
where in the last step, we changed the integration variable from
$\beta$ to $T=1/(\beta k)$. As mentioned in Section II,
in the thermodynamical definition, an entropy change is given by
$$\Delta S=k\Delta\Sigma=\int \frac{\mbox{d}Q}{T}$$
along a reversible process, but $\mbox{d}Q=C(T)\mbox{d}T$,
where $C(T)$ is the heat capacity (at constant volume), and so,
$$\Delta S=
\int_{T_0}^{T_1}\frac{\mbox{d}TC(T)}{T}.$$
Thus, we identify the heat capacity
pertaining the Gaussian broadcast channel as
$C(T)=k/2$, independently of $T$, 
which is exactly the same as the heat capacity (per degree of freedom) 
of an ideal gas without gravitation
(cf.\ e.g., \cite[Sect.\ 4.4, p.\ 106]{Kardar07}).\footnote{The classical
heat capacity per particle 
of an ideal gas at constant volume is actually $C=3k/2$.
The extra factor of $3$ accounts for three degrees of freedom
per particle, owing to the three dimensions of space.}
This is because
the Gaussian channel, considered
in this example, induces a quadratic Hamiltonian, just like
that of the ideal gas (cf.\ the first term
of eq.\ (\ref{idealgas})). 

It is instructive to examine also the case where the directions of the
additive channels are reversed, or equivalently, to examine
the difference $I(U;V)-I(X;V)$ for the original channels defined above. 
Adopting the latter definition, and using the main results of \cite{GSV05},
concerning the relation between $I(U;V)$ and
the minimum mean square error (MMSE), $\mbox{mmse}(V|U)$, in estimating $V$
from $U$ (and of course, similar relations for $X$ and $V$),
we find that the increase in entropy is:
\begin{eqnarray}
I(U;V)-I(X;V)
&=&\frac{1}{2}\int_0^{\beta_0}
\mbox{mmse}\left(V\bigg|V+\frac{N}{\sqrt{\beta}}\right)\mbox{d}\beta-\nonumber\\
&
&\frac{1}{2}\int_0^{\beta_1}\mbox{mmse}\left(V\bigg|V+
\frac{N}{\sqrt{\beta}}\right)\mbox{d}\beta\nonumber\\
&=&\frac{1}{2}\int_{\beta_1}^{\beta_0}\mbox{mmse}
\left(V\bigg|V+\frac{N}{\sqrt{\beta}}\right)\mbox{d}\beta\nonumber\\
&=&\int_{T_0}^{T_1}\frac{\mbox{mmse}(V|V+N\sqrt{kT})}{2kT^2}\mbox{d}T
\end{eqnarray}
where  $N\sim\calN(0,1)$.
Thus, now we identify the heat capacity as
$$C(T)=\frac{\mbox{mmse}(V|V+N\sqrt{kT})}{2T}.$$
If, in addition, $V$ is zero--mean, Gaussian, with variance $\sigma_V^2$, then
$$C(T)=\frac{k\sigma_V^2}{2(\sigma_V^2+kT)}.$$
In the high--SNR regime ($\sigma_V^2 \gg kT$), this gives $C(T)\approx k/2$,
which is the same as before.\\

\noindent
{\it Example 4 -- binary symmetric degraded broadcast channel:}
In a similar manner, consider the binary symmetric degraded broadcast
channel, that is, a cascade of two binary symmetric channels,
$$X=U\oplus N_1;~~U=V\oplus N_2,$$ 
where all RV's are binary $\{0,1\}$,
$\oplus$ designates addition modulo 2, and $(X,N_1,N_2)$ are independent.
In this case, the Hamiltonian is $\calE(x)=E_0x$, $x\in\{0,1\}$, where
$E_0$ is a constant (having the units of energy), and
we have 
$$\mbox{Pr}\{N_1=x\}=\frac{e^{-\beta_0E_0x}}{1+e^{-\beta_0E_0x}}~~x\in\{0,1\}$$ 
and similarly,
$$\mbox{Pr}\{N_1\oplus N_2=x\}=\frac{e^{-\beta_1E_0x}}{1+e^{-\beta_1E_0x}}.$$ 
Here the
heat capacity can be shown to be given by:
$$C(T)=\frac{E_0^2}{kT^2}\cdot\frac{e^{-E_0/(kT)}}{[1+e^{-E_0/(kT)}]^2},$$
which agrees with the heat capacity of 
a system of two--level non--interacting particles
(see, e.g.\ \cite[Sect.\ 4.3, eq.\ (4.22)]{Kardar07}).

\section{Error Exponents and Reversible Processes}

We mentioned the notion of a reversible process, and the question that
might naturally arise, at this point, concerns the information--theoretic
analogue of this term.
This seems to have a direct relationship to the behavior of error exponents of hypothesis
testing
and the Neyman--Pearson lemma: Let $P_0(x)$ and $P_1(x)$ be two probability
distributions (or densities, in the continuous case) of a random variable $X$,
taking values in
an alphabet $\calX$. Given an observation $x\in\calX$, one would
like to decide whether it emerged from $P_0$ or $P_1$. A decision rule is a
partition of $\calX$ into two complementary regions $\calX_0$ and $\calX_1$,
such that whenever $X\in\calX_i$ one decides 
in favor of the hypothesis that
$X$ has emerged from $P_i$, $i=0,1$. Associated with any decision rule, there are
two kinds of error probabilities: $P_0(\calX_1)$ is the probability of
deciding in favor of $P_1$ while $x$ has actually generated by $P_0$, and
$P_1(\calX_0)$ is the opposite kind of error. The Neyman--Pearson
problem is about the quest for the optimum decision
rule in the sense of minimizing $P_1(\calX_0)$ subject to the constraint that
$P_0(\calX_1)\le \alpha$ for a prescribed constant $\alpha\in[0,1]$.
The Neyman--Pearson lemma asserts that the optimum decision rule, in this
sense, is given by the likelihood ratio test (LRT)
$\calX_0^*=(\calX_1^*)^c=\{x:~P_0(x)/P_1(x)\ge \mu\}$, where
the threshold $\mu=\mu(\alpha)$ is tuned so as to meet the constraint
$P_0(\calX_1)\le \alpha$
with equality (assuming that this is possible). 

Assume now that instead of one
observation $x$, we
have a vector $\bx$ of $n$ i.i.d.\
observations $(x_1,\ldots,x_n)$, emerging either all from $P_0$, or all from
$P_1$. In this case, the error probabilities of the two kinds,
pertaining to the LRT, $P_0(\bx)/P_1(\bx)\ge \alpha$, can decay
asymptotically exponentially, provided that $\alpha=\alpha_n$ is chosen to
decay exponentially with $n$ (though not too fast), and the asymptotic
exponents,
$e_0=\lim_{n\to\infty}[-\frac{1}{n}\ln P_0(\calX_1^*)]$ and
$e_1=\lim_{n\to\infty}[-\frac{1}{n}\ln P_1(\calX_0^*)]$
can be easily found (e.g., by using the method of types) to be
$$e_i(\lambda)=D(P_\lambda\|P_i)=\sum_{x\in\calX}P_\lambda(x)\ln
\frac{P_\lambda(x)}{P_i(x)}; ~~i=0,1$$
where
$$P_\lambda(x)=\frac{P_0^{1-\lambda}(x)P_1^\lambda(x)}{Z(\lambda)}$$
with
$$Z(\lambda)=\sum_{x\in\calX} P_0^{1-\lambda}(x)P_1^\lambda(x)$$
and $\lambda\in[0,1]$ being a parameter (depending on $\mu$) that controls the
tradeoff between
the error exponents of the two kinds: For $\lambda=0$, $e_0(0)=0$ and
$e_1(0)=D(P_0\|P_1)$. As $\lambda$ grows from $0$ to $1$, $e_0(\lambda)$
increases and $e_1(\lambda)$ decreases. Finally, for $\lambda=1$,
$e_0(1)=D(P_1\|P_0)$ and $e_1(1)=0$.

From the physics point of view, given $P_0$ and $P_1$, let us define
the Hamiltonians, $\calE_0(x)=-\ln P_0(x)$ and
$\calE_1(x)=-\ln P_1(x)$, and let the inverse temperature be set to
$\beta=1$. Let $P_\lambda(x)$ be defined as above, which can be referred to as
the Boltzmann distribution with Hamiltonian
$\calE_\lambda(x)=(1-\lambda)\calE_0(x)+\lambda\calE_1(x)$ and $\beta=1$.
Let $\lambda_t$, $t\in[0,\tau]$, be a function that starts from
$\lambda_0=0$ and ends at $\lambda_\tau=1$. Now,
assuming that the conditions for the Jarzynsky equality hold
in this case, the average work along
the process, which is
$$\bE\{W\}=\int_0^\tau \mbox{d}\lambda_t\cdot\bE_{\lambda_t}\{
\calE_1(X)-\calE_0(X)\},$$
cannot be smaller than $\Delta F$, which in this case vanishes.
As said, equality $\bE\{W\}=\Delta F\equiv 0$ is attained for
a reversible process.

Indeed, these relations can easily be seen to hold here and also be related to the
error exponents of Neyman--Pearson testing, and even from a direct derivation,
without recourse to physical considerations: Considering the Hamiltonians
$\calE_i(x)=-\ln P_i(x)$, $i=0,1,$ as mentioned above,
we have:
\begin{eqnarray}
\bE\{W\}&=&\int_0^\tau
\mbox{d}\lambda_t\bE_{\lambda_t}\ln\frac{P_0(X)}{P_1(X)}\nonumber\\
&=&\int_0^\tau
\mbox{d}\lambda_t\sum_{x\in\calX}P_{\lambda_t}(x)\ln\frac{P_0(x)}{P_1(x)}\nonumber\\
&=&\int_0^\tau
\mbox{d}\lambda_t[D(P_{\lambda_t}\|P_1)-D(P_{\lambda_t}\|P_0)]\nonumber\\
&=&\int_0^\tau
\mbox{d}\lambda_t[e_1(\lambda_t)-e_0(\lambda_t)]
\end{eqnarray}
On the other hand, we can also rewrite the second line of the last chain of equalities
as:
\begin{equation}
\bE\{W\}=-\int_0^\tau\mbox{d}\lambda_t\cdot\left[\frac{\partial\ln
Z(\lambda)}{\partial\lambda}\right]_{\lambda=\lambda_t}.
\end{equation}
Now, if $\{\lambda_t\}$ is everywhere differentiable (which is analogue
to a reverisble process), this amounts to
\begin{eqnarray}
\bE\{W\}&=-&\int_0^\tau\mbox{d}t\dot{\lambda}_t\cdot\left[\frac{\partial\ln
Z(\lambda)}{\partial\lambda}\right]_{\lambda=\lambda_t}\nonumber\\
&=&-\int_0^\tau\mbox{d}t\cdot\frac{\mbox{d}\ln Z(\lambda_t)}{\mbox{d}
t}\nonumber\\
&=&\ln Z(\lambda_0)-\ln Z(\lambda_\tau)\nonumber\\
&=&\ln Z(0)-\ln Z(1)\nonumber\\
&=&\ln 1-\ln 1=0.
\end{eqnarray}
If, on the other hand, $\{\lambda_t\}$ contains jump--discontinuities, 
then every such jump, say, from $\lambda_1$ to $\lambda_2$, contributes
to the integral a term of the form
$$\mbox{d}\lambda_t\cdot\left[\frac{\partial\ln Z(\lambda)}{\partial\lambda}\right]_{\lambda=\lambda_t}
=(\lambda_2-\lambda_1)\cdot\left[\frac{\partial\ln
Z(\lambda)}{\partial\lambda}\right]_{\lambda=\lambda_1},$$
which is smaller than $\ln Z(\lambda_2)-\ln Z(\lambda_1)$, due to the convexity
of the function $\ln Z(\lambda)$. Consequently, because of the minus sign, each such
discontiuity increases $E\{\bW\}$ above zero.
Thus, we indeed see that,
$$\int_0^\tau \mbox{d}\lambda_te_1(\lambda_t)\ge 
\int_0^\tau \mbox{d}\lambda_te_0(\lambda_t)$$
with equality in the differentiable (reversible) case. 
This in turn means that in this case,
$$\int_0^\tau \mbox{d}t\dot{\lambda}_te_0(\lambda_t)=
\int_0^\tau \mbox{d}t\dot{\lambda}_te_1(\lambda_t).$$
The left-- (resp.\ right--) hand side is simply $\int_0^1\mbox{d}\lambda
e_0(\lambda)$ (resp.\ $\int_0^1\mbox{d}\lambda e_1(\lambda)$)
which means that the areas under the graphs of the functions $e_0$ and $e_1$
are always the same. 

While these integral relations between the error exponent functions have
actually been derived without recourse to any physical considerations,
it is the physical point
of view that gives the trigger to point out these relations.

\section*{Acknowledgement}
The author thanks Shlomo Shamai for the suggesting the problem,
as well as Yariv Kafri and Dov Levine for useful discussions
and for bringing ref.\ \cite{PKL08} to his attention.

\end{document}